\documentclass[11pt,a4paper]{article}
\usepackage{jcappub}
\usepackage{graphicx}
\usepackage{dcolumn}
\usepackage{bm}
\usepackage{booktabs}
\usepackage{colortbl}
\usepackage{collcell}
\usepackage{enumerate}
\usepackage{float}
\usepackage{verbatim}
\usepackage{multirow}
\usepackage{xparse}

\definecolor{rossos}{cmyk}{0,1,1,0.55}
\definecolor{mygreen}{rgb}{0.27, 0.64, 0.48}
\definecolor{mygray}{gray}{.95}

\newcommand{\virg}[1]{``#1''}

\def\min{\text{min}}
\def\max{\text{max}}

\def\DM{\text{DM}}

\def\CM{\text{c.m.}}
\def\LOS{\text{LOS}}
\def\BH{\text{BH}}
\def\doppler{\mathcal{D}}

\title{Blazar-Boosted Dark Matter at Super-Kamiokande}

\author{Alessandro Granelli,} 
\author{Piero Ullio and}
\author{Jin-Wei Wang}

\affiliation{Scuola Internazionale Superiore di Studi Avanzati (SISSA),\\via Bonomea 265, 34136 Trieste, Italy}
\affiliation{Istituto Nazionale di Fisica Nucleare (INFN), Sezione di Trieste,\\via Valerio 2, 34127 Trieste, Italy}
\affiliation{Institute for Fundamental Physics of the Universe (IFPU),\\via Beirut 2, 34151 Trieste, Italy}

\emailAdd{agranell@sissa.it}
\emailAdd{ullio@sissa.it}
\emailAdd{jinwwang@sissa.it (corresponding author)}

\abstract{
Dark matter particles near the center of a blazar, after being accelerated by the elastic collisions with relativistic electrons and protons in the blazar jet, can be energetic enough to trigger detectable signals at terrestrial detectors.
In this work, focusing on the blazars TXS 0506+056 and BL Lacertae, we derive novel limits on the cross section of the elastic scattering between dark matter and electrons by means of the available Super-Kamiokande data.
Thanks to the large blazar-boosted dark matter flux, the limit on the dark matter-electron scattering cross section for dark matter masses below 100 MeV can be as low as $\sim10^{-38}~\text{cm}^2$, which is
orders of magnitude stronger than the analogous results from galactic cosmic rays.}

\begin{document}

${}$\vskip 3cm
\vspace*{-15mm}
\begin{flushright}
SISSA 02/2022/FISI
\end{flushright}
\vspace*{0.7cm}

\maketitle

\section{Introduction}
\label{sec:intro}
The existence of dark matter (DM) has been established by solid astrophysical and cosmological observations, while its particle nature is still unknown \cite{hep-ph/0404175, 1807.06209}. Assuming that DM has incredibly feeble interaction with ordinary matter, direct DM detection experiments (e.g.~XENON1T \cite{1206.6288, XENON:2017lvq}, PandaX-II \cite{PhysRevLett.119.181302}) are promising in detecting DM scatterings with target nuclei. However, for DM masses below $\sim$ 1 GeV,
the typical kinetic energy of DM in the local halo is not enough to imprint a recoil energy above the threshold of $\sim$ 1 keV, thus leading to a rapid decrease in detection sensitivity.
Nevertheless, DM in this mass region may still be energetic enough to trigger detectable electronic recoils thanks to a lower threshold ($\sim0.186$ keV \cite{XENON:2019gfn}), thus opening an alternative way to explore sub-GeV DM \cite{PandaX-II:2021nsg,XENON:2019gfn}. Still, the search for DM interacting with electrons in direct detection setups inevitably faces the same problem for DM masses below $\sim 10$ MeV \cite{PandaX-II:2021nsg,XENON:2019gfn}.

In the past few years, to circumvent the limitations of light DM direct detection, some scenarios with \virg{boosted} DM populations have been proposed \cite{1405.7370,1506.04316, 1708.03642, 1709.06573}. For instance, in Ref.~\cite{PhysRevLett.122.171801} the authors put forward a novel idea for which DM particles in the local halo are accelerated via elastic collisions with galactic high-energy cosmic rays (CRs). This small but inevitable component of DM (dubbed CRDM) possesses enough energy to set constraints on the DM-proton interaction cross section ($\sigma_{\chi p}$) for sub-GeV DM. Another similar study on CRDM-electron scattering at Super-Kamiokande (Super-K) can be found in Refs.~\cite{Super-Kamiokande:2017dch,PhysRevLett.122.181802, PhysRevD.100.103011}, where stringent constraints on DM-electron scattering cross section ($\sigma_{\chi e}$) for DM masses down to $\sim$ 1 keV are presented ($\sigma_{\chi e}\lesssim 10^{-33} ~\text{cm}^{2}$). 

Recently, a new DM acceleration mechanism at blazars, referred to as Blazar-Boosted DM (BBDM), has been suggested \cite{wang2021direct}. Through the scatterings with high-energy protons in the jet of a blazar, DM particles can be boosted up to high velocities. Moreover, the existence of a supermassive Black Hole (BH) at the blazar center may provide a dense DM population \cite{PhysRevLett.83.1719}. The combination of these two distinguished characteristics makes blazars ideal DM boosters, that can induce a DM flux at Earth much stronger than that from galactic CRs. 
In Ref.~\cite{wang2021direct} the authors focused on the blazars TXS 0506+056 and BL Lacertae and derived the corresponding constraints on $\sigma_{\chi p}$ from the results of direct DM detectors (e.g.~XENON1T \cite{XENON:2017lvq}), as well as neutrino detectors (e.g.~MiniBooNE \cite{MiniBooNE:2008paa} and Borexino \cite{Borexino:2000uvj}). 

For the sake of simplicity, the authors of Ref.~\cite{wang2021direct} ignored the scattering between DM and electrons. However, it is actually intriguing to analyse the influence of DM-electron scattering for the following reasons. Firstly, the observations of the photon Spectral Energy Distribution (SED) of blazars clearly reveal two peaks, one in the infrared/X-ray bands and the other at $\gamma$-ray frequencies \cite{0912.2040}. Different SED models agree that the low-energy peak is due to the synchrotron emission of electrons \cite{1992ApJ...397L...5M, 1993A&A...269...67M, 1996ApJ...461..657B, astro-ph/0004052, 0711.4112, 0909.0932, 1304.0605}, suggesting that the electron component of the blazar jets is essential (see, e.g.,~\cite{2012.13302} for a recent SED model review). Secondly, it provides a possible way to study the characteristics of blazar jet models and the nature of DM by searching for DM-electron recoil signals at ground detectors. Based on Ref.~\cite{wang2021direct}, in this work we investigate the framework of BBDM from the blazars TXS 0506+056 and BL Lacertae. We activate the effects of $\sigma_{\chi e}$ and calculate the corresponding constraints from Super-K observations. 
In particular, since the flux of BBDM can be much larger and extend to much higher energies than that of CRDM \cite{wang2021direct}, we expect that more stringent constraints on $\sigma_{\chi e}$ can be derived from
Super-K results \cite{Super-Kamiokande:2017dch}. 

This work is organized as follows. In Sec.~\ref{sec:jet} we compute the jet spectrum for the two blazars under consideration. In Sec.~\ref{sec:BBDM_flux}, we calculate the DM density profile and estimate the BBDM flux. We devote Sec.~\ref{sec:DD} to the computation of the constraints at Super-K and summarise our results in Sec.~\ref{sec:Conclusion}.

\section{Blazar Jet Spectrum}
\label{sec:jet}

The jets of blazars can be well described by the \virg{blob geometry} \cite{Dermer2009-DERHER}: electrons and protons move isotropically in the blob frame with a power-law energy distribution,
and, in the BH center-of-mass rest frame (also observer's frame), the blob itself moves along the jet axis with speed $\beta_B$. The corresponding Lorentz boost factor reads $\Gamma_B \equiv (1-\beta_B^2)^{-1/2}$.
The misalignment angle between the jet axis and the observer's line-of-sight, hereafter denoted by $\theta_\LOS$, is usually of few degrees. 
The desired jet spectrum in the observer's frame can be derived from a Lorentz boost transformation and can be expressed as (see Ref.~\cite{wang2021direct} for a detailed derivation) 
\begin{equation}
\label{eq:CRSpectrum}
    \frac{d\Gamma_j}{dT_jd\Omega} 
    = \frac{1}{4\pi}c_j\,\left(1+\frac{T_j}{m_j}\right)^{-\alpha_j} \frac{\beta_j(1-\beta_j\beta_B  \mu)^{-\alpha_j} \Gamma_B^{-\alpha_j}}{\sqrt{(1-\beta_j \beta_B \mu)^2 - (1-\beta_j^2)(1-\beta_B^2)}}\,,
\end{equation}
where the subscript $j \in \{e,\,p\}$ refers either to electrons or protons with masses $m_e \simeq 0.511$ MeV and $m_p \simeq 0.938$ GeV, respectively, 
$\alpha_j$ is the spectral power index, $T_{j}$ and $\beta_j = \left[1-m_j^2/(T_j+m_j)^2\right]^{1/2}$ are respectively the kinetic energy and speed of the particle,  $c_j$ is the normalisation constant that can be computed from the luminosity $L_j$ (see further in Eq.~\eqref{eq:luminosity} and \eqref{eq:luminosity2}), $\mu$ is the cosine of the angle between the particle's direction of motion and the jet axis.

The relevant (Lepto-)Hadronic SED model parameters of TXS 0506+056 \cite{TXS, TXS_2} and BL Lacertae \cite{1304.0605} are summarized in Table \ref{Tab:HadronicModel}.
The quantities $\gamma'_{\min,\,j}$ and $\gamma'_{\max,\,j}$ are the minimal and maximal Lorentz boost factors in the blob frame, while $\doppler = [\Gamma_B\left(1-\beta_B\cos\theta_\LOS\right)]^{-1}$ is the Doppler factor. 
In practice, two different assumptions are commonly used in the blazar jet model fitting, $\doppler = \Gamma_B$ or $\doppler = 2\Gamma_B$.
For the first assumption, $\theta_\LOS$ can be solved by using the definition of $\doppler$, while for the second assumption, $\theta_\LOS$ is set to zero. 
In the same table, the other relevant parameters of the considered blazars are also given, including the redshift $z$ \cite{BLRedshift, 1802.01939}, the luminosity distance $d_L$, and central BH mass $M_\BH$ \cite{Titarchuk:2017jwu, 1901.06998}, together with the specific values of the normalisation constant $c_j$ appearing in Eq.~\eqref{eq:CRSpectrum}, that can be fixed via the relation \cite{PhysRevD.82.083514, wang2021direct}
\begin{equation}
    L_j = \int d\Omega \int dT_j \left(T_j+m_j\right) \frac{d\Gamma_j}{dT_j d\Omega} = c_j m_j^2 \Gamma_B^2 \int_{\gamma'_{\min,\,j}}^{\gamma'_{\max,\,j}}d\gamma'_j\, (\gamma'_j)^{1-\alpha_j}\,,
    \label{eq:luminosity}
\end{equation}
giving
\begin{equation}
    c_j = \frac{L_j}{m_j^2\Gamma_B^2}\times \begin{cases} (2-\alpha_j)/\left[(\gamma'_{\max,\,j})^{2-\alpha_j}-(\gamma'_{\min,\,j})^{2-\alpha_j}\right]\,& \text{if } \alpha_j\neq2\,;\\
    1/\log{\left(\gamma'_{\max,\,j}/\gamma'_{\min,\,j}\right)}\,&\text{if } \alpha_j=2\,.
    \end{cases}
    \label{eq:luminosity2}
\end{equation}
We note that $\gamma_{\min,\,e}'\gg\Gamma_B$ for both sources, namely, the electrons are ultra-relativistic in the blob frame and remain so in the observer's frame. Therefore, we decide to adopt the approximation $\beta_e \approx 1$.

\begin{table}
\begin{center}
\begin{tabular}{ccc}
    \toprule
    \rowcolor[gray]{.95}
    \multicolumn{3}{c}{\bf (Lepto-)Hadronic Model Parameters}\\
    \hline
    \hline
     \rule{0pt}{2.5ex} 
    ~~~Parameter (unit)~~~ & TXS 0506+056~~~ & BL Lacertae~~~ \\
    \hline
    \rule{0pt}{2.5ex} 
     $z$   &0.337 & 0.069 \\
     $d_L$ (Mpc)    & 1835.4 & 322.7 \\
     $M_\BH$ ($M_\odot$) &$3.09\times10^{8} $ &$8.65\times 10^7$\\
    $\mathcal{D}$  &40$^\star$& 15  \\
    $\Gamma_B$ &20& 15  \\
     $\theta_\LOS\, (^\circ)$  &$0$&$3.82$  \\
    $\alpha_p$   & $2.0$& $2.4$\\
     $\alpha_e$   & $2.0$& $3.5$\\
    $\gamma'_{\min,\,p}$ & 1.0 & 1.0  \\
     $\gamma'_{\max,\,p}$ & $5.5\times10^{7^\star}$& $1.9\times10^9$  \\
     $\gamma'_{\min,\,e}$ & 500 & 700  \\
     $\gamma'_{\max,\,e}$ & $1.3\times10^{4^\star}$& $1.5\times10^4$  \\
     $L_p$ (erg/s)  & $2.55 \times 10^{48^\star}$& $9.8 \times 10^{48}$  \\
     $L_e$ (erg/s)  & $1.32 \times 10^{44^\star}$& $8.7 \times 10^{42}$  \\
     $c_p$ ($\text{s}^{-1}\text{sr}^{-1}\text{GeV}^{-1}$)  & $2.54 \times 10^{47}$& $1.24 \times 10^{49}$\\
     $c_e$ ($\text{s}^{-1}\text{sr}^{-1}\text{GeV}^{-1}$)  & $2.42 \times 10^{50}$& $2.59 \times 10^{54}$\\
    \hline
    \hline
\end{tabular}
\caption{The model parameters for the blazars TXS 0506+056 (Lepto-Hadronic) \cite{TXS, TXS_2} and BL Lacertae (Hadronic) \cite{1304.0605} used in our calculations.
The quantities flagged with a star ($^\star$) correspond to mean values computed from the ranges given in the second column of Table 1 of Ref.~\cite{TXS_2}.\textsuperscript{a}
In the model fitting, the assumption of $\doppler = 2\Gamma_B$ ($\Gamma_B$) is used for TXS 0506+056 (BL Lacertae). The resulting values of the normalisation constants $c_{e,\,p}$, as well as the redshift $z$ \cite{BLRedshift,1802.01939}, luminosity distance $d_L$, and BH mass $M_\BH$ \cite{Titarchuk:2017jwu, 1901.06998} for the two considered sources are also reported.}
\label{Tab:HadronicModel}
\end{center}
\footnotesize\textsuperscript{a} Considering that these two sample blazars are BL Lac-type, which means their luminosity is time-dependent, we will briefly discuss how this can affect our final results in Sec.~\ref{sec:resB}.
\end{table}

\section{Dark Matter Profile and Flux from Blazars}
\label{sec:BBDM_flux}

\subsection{Dark Matter Density Profile}
The adiabatic growth of a BH in the central region of a DM halo is expected to focus the distribution of DM particles, giving rise to a very dense spike. The idea was originally suggested by Gondolo and Silk~\cite{PhysRevLett.83.1719} discussing the case for the BH at the center of the Milky Way (MW). Implementing angular momentum and radial action as adiabatic invariants, it was shown that an ergodic, self-gravitating, single power-law spherical DM profile $\rho(r) \propto r^{-\gamma}$ is turned into a sensibly steeper profile~\cite{PhysRevLett.83.1719}:
\begin{equation}\label{eq:rho_prime}
   \rho'(r)  \propto  r^{-\alpha}\quad \text{with } \alpha = \frac{9-2\gamma}{4-\gamma}\,.
\end{equation}
Such process has, in particular, dramatic phenomenological implications for DM that can annihilate in pairs into Standard Model particles, since it would imply that the region around the MW BH is an extremely bright source for indirect DM detection signals. The effect can be so large that the pair annihilation itself may deplete the central DM density: considering this effect as a continuous loss since the time of the BH formation, and ignoring any replenishing from the surrounding environment, there is a maximum surviving DM density $\rho_\text{core} \simeq m_\chi/(\left\langle \sigma v \right \rangle_0 t_\BH)$, with $\left\langle \sigma v \right \rangle_0$ being the DM annihilation cross section times relative velocity and $t_\BH$ the BH lifetime. The DM profile would then take the final form~\cite{PhysRevLett.83.1719}
\begin{equation} \label{eq:rho_dep}
    \rho_\text{DM}(r) = \frac{\rho'(r) \rho_\text{core}}{\rho'(r) + \rho_\text{core}}\,.
\end{equation}
Given its profound impact on thermal Weakly Interacting Massive Particles (WIMPs), the MW BH spike has been very closely scrutinized, discussing both the validity and implications of the assumptions leading to the spike formation, as well as effects possibly impacting on the spike after the formation. For instance, in Ref.~\cite{Ullio:2001fb} it was shown that, if the dynamical time for DM particles is not much shorter than the BH formation time, going away from the adiabatic growth assumption to the opposite limit of BH appearing instantaneously, then the DM profile is much less focused. With an initial Navarro-Frenk-White (NFW) profile \cite{NFW_profile}, for which $\gamma=1$, this results in a profile with $\alpha = 4/3$ rather than one with the slope $\alpha = 7/3$ has predicted in Eq.~\eqref{eq:rho_prime}. Spherical symmetry is also crucial in the process, and its violation, e.g., by hierarchical mergers onto the central BH may lead to depletion of the central spike~\cite{Ullio:2001fb,PhysRevLett.88.191301} down to a weak power $\alpha = 1/2$. The authors of Ref.~\cite{PhysRevLett.93.061302} argued instead that the presence of an inner stellar cluster around the MW BH would relax, independently of the initial conditions, the DM spike into a \virg{minicusp} with $\alpha = 3/2$; this effect was included in the numerical model of Ref.~\cite{Bertone:2005hw} which resulted in a less severe spike depletion.

In the cases under study, we are actually not in the position of describing the DM spikes around the BHs at the center of the considered blazars in terms of initial conditions prior the BHs' formations and up to their present configurations. We are forced to refer to a simplified model, encompassing however uncertainties mentioned above. In most of our results we will consider an initial NFW profile as modified within the Gondolo and Silk scenario according to Eq.~\eqref{eq:rho_prime}; to fix the normalisation condition for $\rho'(r)$, as in Ref.~\cite{wang2021direct}, 
we consider the region within which the BH is expected to dominate the potential well even after the spike formation and set (in analogy to results in Ref.~\cite{Ullio:2001fb})
\begin{equation}\label{eq:DM_condition}
\int_{4R_S}^{10^5 R_S} 4\pi r^2 \rho'(r) dr \simeq M_\BH\,,
\end{equation}
where $R_S$ is the Schwarzschild radius of the central BH; DM particles within $4R_S$ are captured by the BH, while $10^5 R_S$ is also the typical radius relevant for BH mass estimations \cite{PhysRevD.82.083514}. At the same time, we include a DM depletion effect, which, for simplicity, is parameterised in terms of a DM pair annihilation rate and the expression in Eq.~\eqref{eq:rho_dep}. We will refer to three benchmark points (BMPs) \cite{wang2021direct}:
\begin{enumerate}[BMP1)]
    \item $\left\langle \sigma v \right \rangle_0 = 0$, so that $\rho_\text{core}\to +\infty$ and $\rho_\DM = \rho'$;
    \item  $\left\langle \sigma v \right \rangle_0 = 10^{-28} \,\text{cm}^3 \,\text{s}^{-1}$ and $t_\BH = 10^9$ yr;
    \item  $\left\langle \sigma v \right \rangle_0 = 3\times10^{-26} \,\text{cm}^3 \,\text{s}^{-1}$ and $t_\BH = 10^9$ yr.
\end{enumerate}
The third benchmark with $\left\langle\sigma v\right \rangle_0 = 3\times 10^{-26}\,\text{cm}^3\,\text{s}^{-1}$ corresponds nominally to the case of DM thermal relics, a regime however which is rather unlikely for the DM masses and interaction strengths considered in our analysis; it also drives, for most of the DM masses of our interest, to a DM core profile $\rho_\text{core}$ extending to larger radii with respect to the inner region considered in Eq.~(\ref{eq:DM_condition}). The case with $\left\langle \sigma v \right \rangle_0=0$ would be appropriate, e.g., for asymmetric DM models \cite{Kaplan:2009ag, Petraki:2013wwa, Zurek:2013wia}, and corresponds more in general to scenarios in which no significant spike depletion effects are expected. The second benchmark stands in between.

The relevant quantity to derive the flux of BBDM is the DM line-of-sight integral $\Sigma_\DM$ (see Sec.~\ref{sec:BBDMflux}), which is defined as \cite{wang2021direct}
\begin{equation}\label{eq:deltaDM}
    \Sigma_\DM(r) \equiv \int_{4 R_S}^{r}  \rho_\DM(r')\,dr'.
\end{equation}
Given that $\Sigma_\DM (r)$ tends to a constant value for $r\gtrsim 10$ pc \cite{wang2021direct} for the mass range of interest, we can factor the effects of the DM profile into $\Sigma_\DM^\text{tot}\equiv \Sigma_\DM(r\simeq 10\,\text{pc})$ \footnote{In case of the formation of a DM spike according to the Gondolo and Silk scenario, the predominate contribution to $\Sigma_\DM(r)$ comes from the very central region, and $\Sigma_\DM^\text{tot}$ is not sensitive to the precise radius at which the DM spike matches onto the galactic halo DM density profile. Such radius may play a role in scenarios is which the spike is reshaped into a much shallower profile after its formation. However, in the examples at hand and for the particle DM mass range of interest, it is unlikely to be as small as few pc. Hence, our choice for the upper extreme of integration, i.e.~10 pc, in $\Sigma_\DM^\text{tot}$ corresponds to a conservative result.}.
\begin{figure}
\centering
\includegraphics[width=0.48\textwidth]{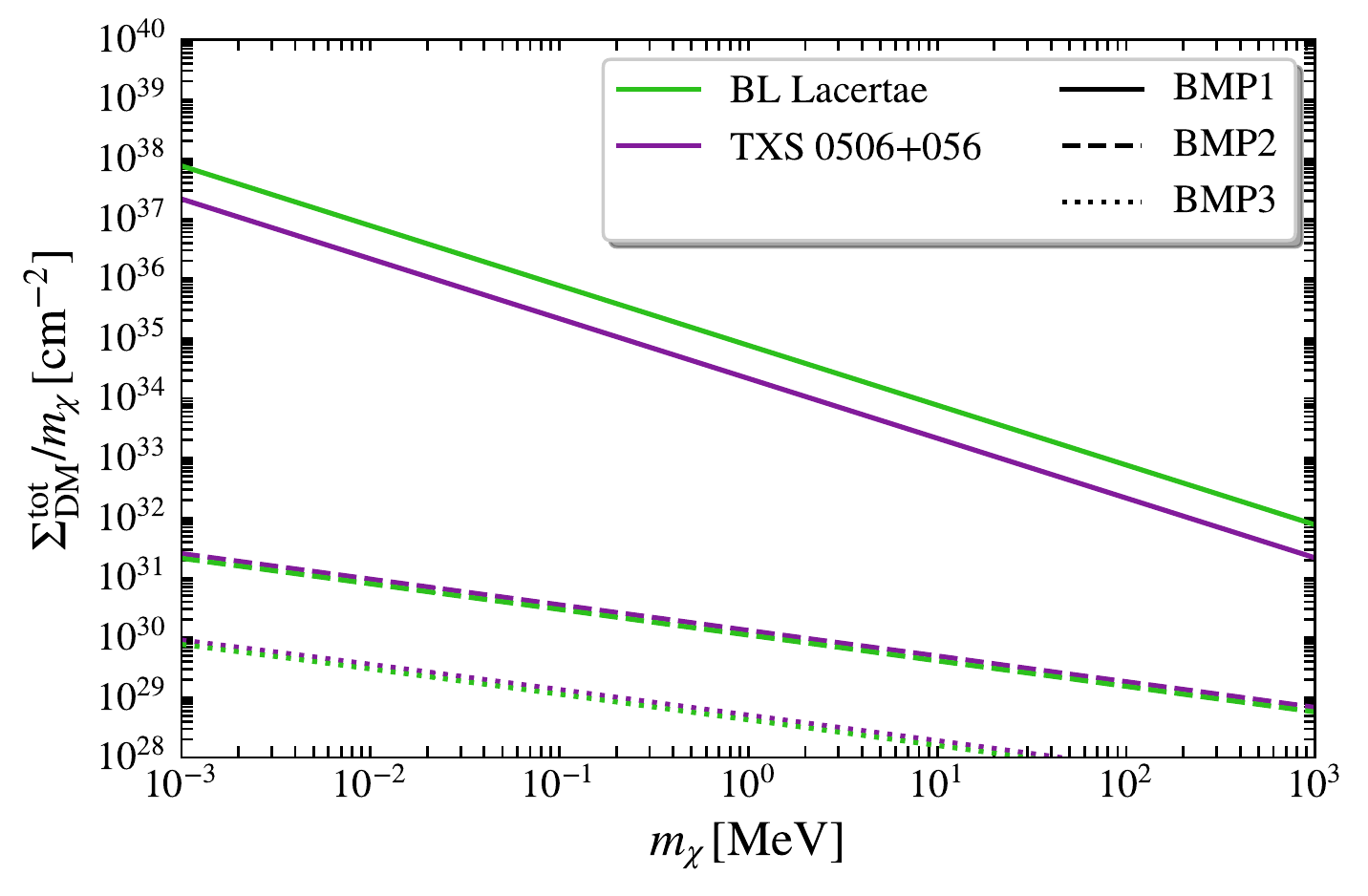}
\includegraphics[width=0.48\textwidth]{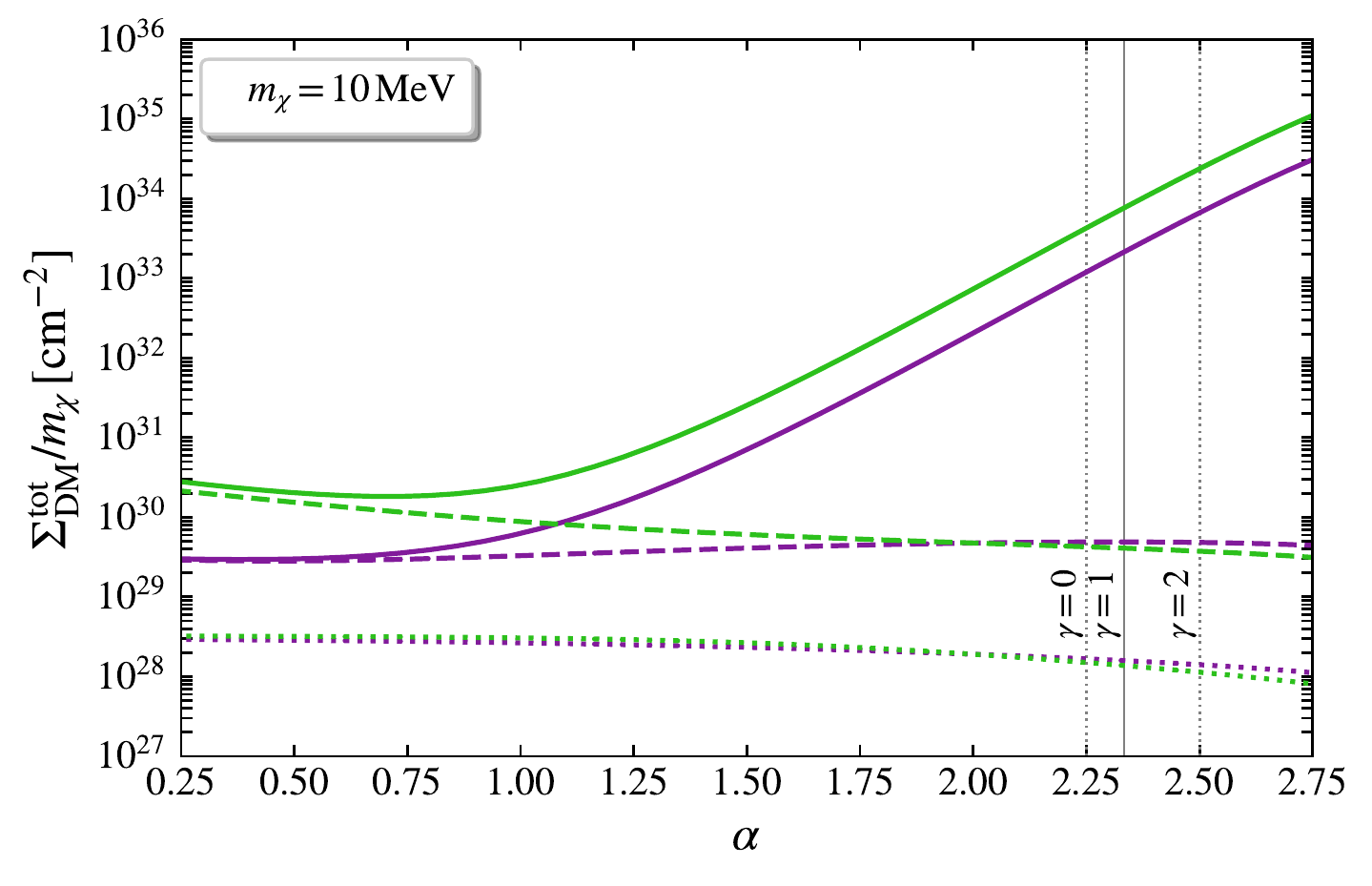}
\caption{The quantity $\Sigma_\DM^\text{tot}/m_\chi$ as a function of $m_\chi$ (left panel) and the slope $\alpha$ (right panel) for BL Lacertae (green) and TXS 0506+056 (purple). The solid, dashed and dotted curves correspond to BMP1, BMP2 and BMP3, respectively. Note that the results shown in the right panel are obtained for $m_\chi = 10$ MeV. The vertical grey lines correspond, from left to right, to the cases of Gondolo and Silk spikes from initial profiles with $\gamma = 0,\,1$ (NFW) and $2$, see Eq.~\eqref{eq:rho_prime}.
}\label{fig:sigmaDM}
\end{figure}
In the left panel of Fig.~\ref{fig:sigmaDM} we show the behaviour of $\Sigma_\DM^\text{tot}/m_\chi$ against $m_\chi$ for TXS 0506+056 (purple) and BL Lacertae (green). For BMP1 (solid lines), $\Sigma_\DM^\text{tot}$ is determined by the properties of the BH ($\rho'(r)$) and independent of $m_\chi$, while for BMP2 (dashed lines) and BMP3 (dotted lines) the dominant contribution to $\Sigma_\DM^\text{tot}$ comes from the core region where $\rho_\text{core}\propto m_\chi$. In the right panel of Fig.~\ref{fig:sigmaDM}, we illustrate the relationship between $\Sigma_\DM^\text{tot}/m_\chi$ and $\alpha$ for different BMPs, setting $m_\chi=10$ MeV (for different masses the plot is scaled).
The solid vertical grey line marks the considered case of a Gondolo and Silk spike with slope $\alpha = 7/3$ (i.e.~$\gamma=1$). It is evident that a smaller choice of $\alpha$ corresponds to an intermediate situation between the already considered cases. In the same plot we also depict how departures from an initial NFW profile, in connection to baryonic feedback models (see, e.g.,~\cite{Martizzi_2013, Cintio_2014, Sameie_2021}), would change the Gondolo and Silk scenario; two different initial conditions with $\gamma = 0$ ($\alpha = 9/4$) and $\gamma = 2$ ($\alpha = 5/2$) are marked by dotted vertical lines. It is clear from the figure that the variation of the initial slope in the range $0\leq\gamma\leq 2$ would only mildly affect the quantity $\Sigma_\DM^\text{tot}/m_\chi$ for BMP1 by factors of few, while leaving the BMP2 and BMP3 cases roughly invariant.

\subsection{Blazar-Boosted Dark Matter Flux}
\label{sec:BBDMflux}
Through elastic collisions, the relativistic electrons and protons in the jet of a blazar can speed up the neighbouring DM particles. Assuming an isotropic differential scattering cross section, the BBDM flux at Earth can be expressed as \cite{wang2021direct}
\begin{equation}\label{eq:spectrumDM}
    \frac{d\Phi_\chi}{dT_\chi} =\frac{\Sigma_\DM^\text{tot}}{2\pi m_\chi d_L^2}\sum_{j=e,\,p}\widetilde{\sigma}_{\chi j}
    \int_{0}^{2\pi}\,d\phi_s\int_{T_j^\min(T_\chi,\phi_s)}^{T_j^\max(T_\chi,\phi_s)}\frac{dT_j}{T_\chi^\max(T_j)}\frac{d\Gamma_j}{dT_j d\Omega}\,,
\end{equation}
where we have summed over the contributions from electrons and protons. The angle $\phi_s$ is the azimuth with respect to the line-of-sight, while the quantity $T_\chi^\max$ is the maximal kinetic energy DM can have after the scattering, i.e.~\cite{PhysRevD.99.063004,wang2021direct, PhysRevLett.122.171801}
\begin{equation}
    T_\chi^\max(T_j) =  \frac{T_j^2 + 2m_j T_j}{T_j + (m_j+m_\chi)^2/(2m_\chi)}\,.
    \label{eq:Tjmax}
\end{equation}
For the DM-proton cross section we assume:
\begin{equation}\label{eq:tildesigma}
    \widetilde{\sigma}_{\chi p} = \sigma_{\chi p} G^2(2m_\chi T_\chi/\Lambda_p^2)\,,
\end{equation}
where $\sigma_{\chi p}$ is the zero-momentum transfer DM-proton cross section and the form factor $G(x^2)\equiv 1/(1 + x^2)^2$ accounts for the internal structure of the proton, with $\Lambda_p\simeq 0.77$ GeV \cite{PhysRevLett.122.171801}. There is no need for a form factor in the case of DM interaction with free electrons, thus $\widetilde{\sigma}_{\chi e} \equiv \sigma_{\chi e}$\footnote{Note that in this work we have assumed a constant $\sigma_{\chi j}$ as in previous literature \cite{wang2021direct, PhysRevD.100.103011, PhysRevLett.122.171801}. A more concrete analysis should include the energy dependence effects, but this would be highly model dependent. For example, in Ref.~\cite{Bondarenko_2020} the authors considered CRDM with a scalar mediator, including the full energy dependence in the cross section (see their Eq.~(3.10)). They found that, for heavy mediators (say with masses larger than $m_\chi$), the final constraints on $\sigma_{\chi p}$ tend to be more stringent, which means that the constant cross section assumption corresponds to a conservative estimation, while for light mediators the effects would be the opposite. We have checked that, after adopting the same model of \cite{Bondarenko_2020}, the conclusions for $\sigma_{\chi j}$ in the BBDM scenario are similar. Anyway, a precise selection or construction of an \virg{ad hoc} model is beyond the scope of this work.}. 

If $\theta_\LOS = 0$, the system is symmetric around the line-of-sight and the integration over $\phi_s$ appearing in Eq.~\eqref{eq:CRSpectrum} is trivial. This is the case for the blazar TXS 0506+056 in the considered Lepto-Hadronic SED model (see Table \ref{Tab:HadronicModel}). If instead the jet is inclined with respect to the line-of-sight, the computation is complicated by the geometry, with the interval of integration and the jet spectrum depending, in general, on $\phi_s$. However, in the case of protons the situation is less involved as they can have arbitrary small energy in the blob frame ($\gamma_{\min,\,p}' = 1$). Thus, the lower limit of integration for protons coincides with the minimal kinetic energy required for the scattering, namely
\begin{equation}
    T_p^\min(T_\chi) = \left(\frac{T_\chi}{2}-m_p\right)\left[1
    \pm\sqrt{1+\frac{\left(m_p+m_\chi\right)^2}{\left(T_\chi-2m_p\right)^2}\frac{2T_\chi}{m_\chi}}\right]\,.
    \label{eq:Tpmin}
\end{equation}
where the $+ (-)$ applies for $T_\chi \geq 2m_p$ ($T_\chi<2m_p$). Also, given that $\gamma_{\max,\, p}'\gg 1$ and that the proton spectrum is strongly attenuated at high energies, the integral over $T_p$ in Eq.~\eqref{eq:CRSpectrum} shows only mild dependency on the upper extreme of integration $T_p^\max$, which we therefore set at $10^7$ GeV with no appreciable loss of accuracy. Conversely, the electron case is more subtle and the extremes of integration result as solutions of kinematical constraints combined with other energy availability conditions (refer to Ref.~\cite{wang2021direct} for more details on the kinematics).

\begin{figure}
\centering
\includegraphics[width=0.48\textwidth]{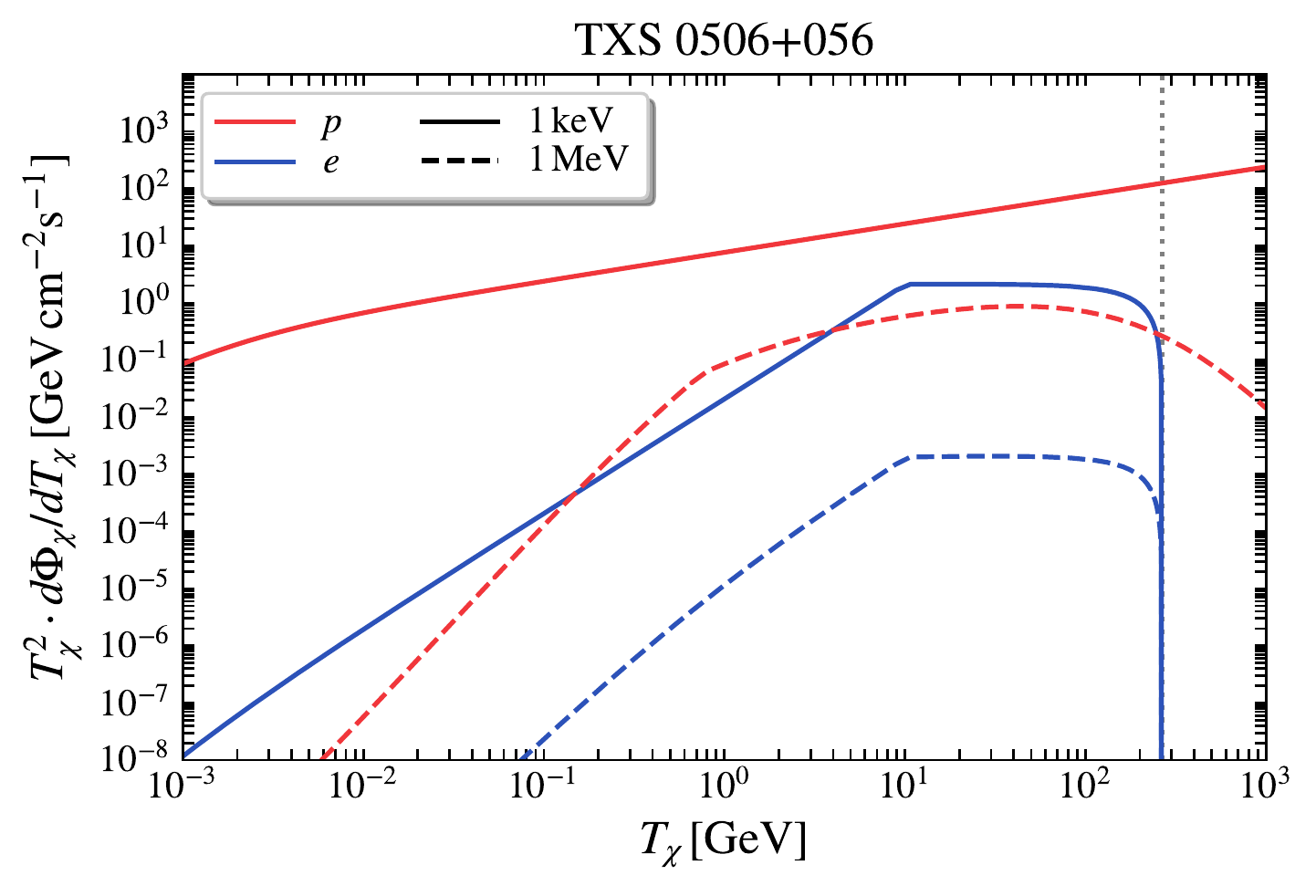}
\includegraphics[width=0.48\textwidth]{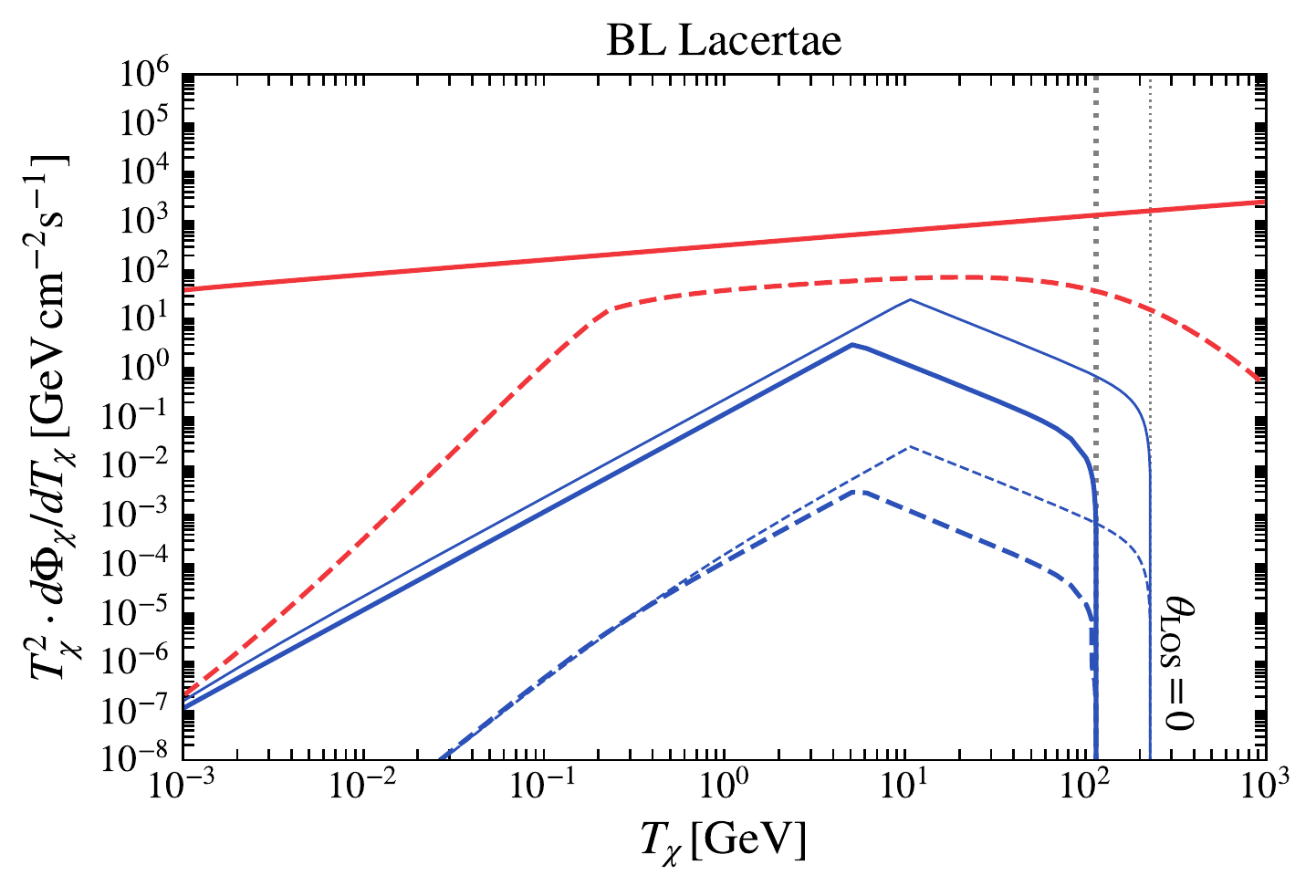}
\caption{The expected flux of BBDM from TXS 0506+056 (left panel) and BL Lacertae (right panel). The red (blue) curves represent the contribution from protons (electrons), while the solid and dashed curves correspond to $m_\chi=1$ keV and 1 MeV, respectively. For BL Lacertae, the thinner blue curves are obtained by artificially setting $\theta_\LOS = 0$. The truncation of the electron-induced BBDM flux is caused by electron energy limitations (see the text for further details). Note that all these results are derived for $\sigma_{\chi p} = \sigma_{\chi e} = 10^{-30}~\text{cm}^2$ and BMP1 parameters.}
\label{fig:DM_Tx_Spectrum}
\end{figure}

We plot in Fig.~\ref{fig:DM_Tx_Spectrum} the flux of BBDM computed numerically for the blazars TXS 0506+056 (left panel) and BL Lacertae (right panel).
The solid and dashed lines are obtained for $m_\chi = 1$ keV and $1$ MeV, respectively. Note that all these results are derived by setting $\sigma_{\chi p} = \sigma_{\chi e} = 10^{-30}~\text{cm}^2$. For clarity, we have separated the contributions from protons (red) and electrons (blue). To underline the effects of the inclination angle, we also show the results of BBDM flux after artificially setting $\theta_\LOS = 0$ for BL Lacertae (thinner lines). In the figure, the BBDM flux from electrons stop at $T_\chi^\max(T_e = \overline{T}_e) \simeq 265.5$ GeV for TXS 0506+056 and 114.9 GeV (229.6 GeV if $\theta_\LOS$ is set to zero) for BL Lacertae, with $\overline{T}_e \simeq m_e[ \gamma_{\max,\,e}'\Gamma_B^{-1}(1-\beta_B\cos\theta_\LOS)^{-1} - 1]$ being the maximal kinetic energy of electrons along the line-of-sight. 
Moreover, it is clear from the plot that the BBDM flux from protons is much larger than that from electrons, even for the same cross section (i.e.~$\sigma_{\chi p} = \sigma_{\chi e}$), due to the fact that $L_p \gg L_e$ in the SED models under consideration.

%%%%%%%%%%%%%%%%%%%%
\section{Constraints on Dark Matter-Electron Scattering Cross Section}
\label{sec:DD}
\subsection{Data Selection and Analysis}
Recently, Super-K has performed a search for CRDM in its \virg{electron elastic scatter-like} event with electron recoil energy $T_e>100$ MeV \cite{Super-Kamiokande:2017dch,PhysRevLett.122.181802, PhysRevD.100.103011}. Due to the large volume (22.5 kt in fiducial volume) and long exposure time (2628.1 days), Super-K is an ideal detector to search for DM-electron scattering signals. In the analysis of Ref.~\cite{Super-Kamiokande:2017dch}, because of the strong energy dependence of the atmospheric neutrino background, three energy bins were considered, namely $0.1<T_e/\text{GeV}<1.33~$ (Bin1), $1.33 <T_e/\text{GeV}<20$ (Bin2), and $20<T_e/\text{GeV}<10^3$ (Bin3). For each bin, the authors of Ref.~\cite{Super-Kamiokande:2017dch} give the total number of data events ($N_\text{Data}$), the Monte Carlo simulated atmospheric neutrino background ($N_\text{Bkg}$) and the signal efficiency ($\epsilon_\text{sig}$), as well as the spatial distribution of the events.

A BBDM particle can hit an electron in the Super-K water tank and imprint a detectable signal. We treat the electrons in the detector as free and at rest in the observer's frame. According to Fig.~\ref{fig:DM_Tx_Spectrum}, the BBDM spectrum can extend to high-energy scales, so that we can expect the scattered electron in the detector to be strongly forward. Moreover, both TXS 0506+056 and BL Lacertae can be treated as point sources. Therefore, by selecting signals from a proper \virg{searching cone} in the direction of the source, we can get rid of most of the background and obtain a higher sensitivity.

\begin{figure}
\centering
\vspace{2em}
    \includegraphics[width=0.46\textwidth]{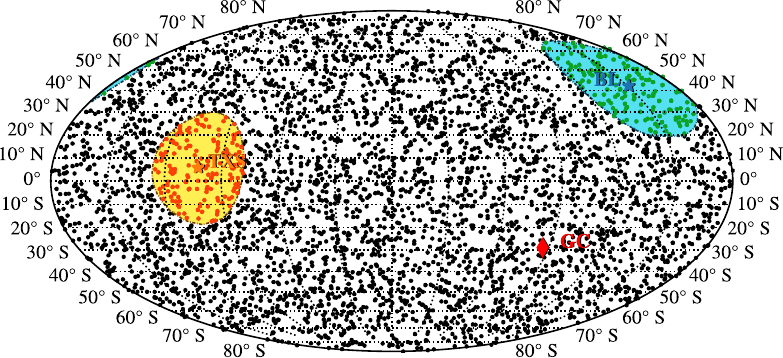}
    \hspace{1em}
    \includegraphics[width=0.46\textwidth]{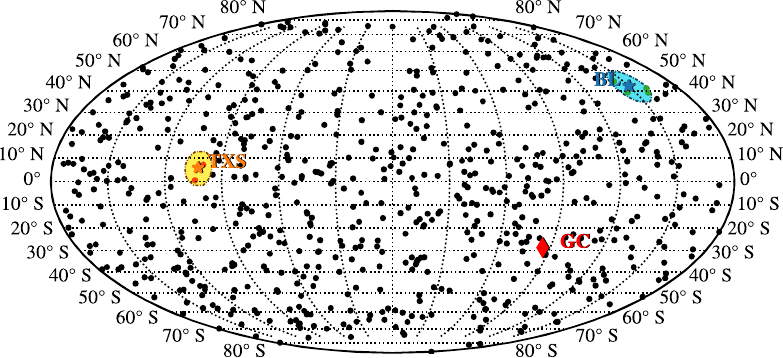}\\
    \vspace{1em}
    \includegraphics[width=0.46\textwidth]{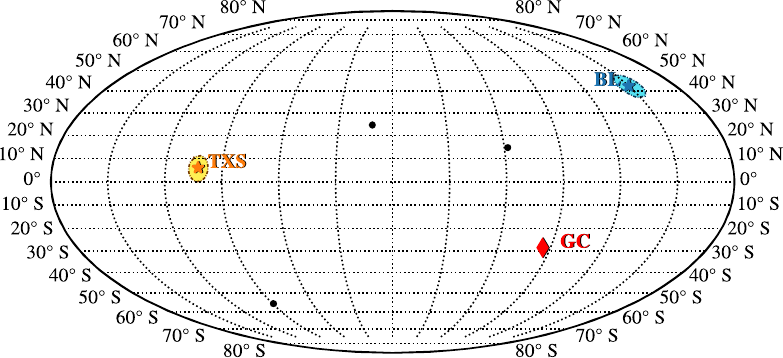}
    \caption{The spatial distribution of the Super-K data events in Bin1 (top-left panel), Bin2 (top-right panel), and Bin3 (bottom panel) \cite{Super-Kamiokande:2017dch}.
    The positions of TXS 0506+056 (TXS) and BL Lacertae (BL) are marked with an orange and a blue star, respectively. As a landmark, the position of the Galactic Center (GC) is also identified (red diamond).
    The yellow (blue) regions around TXS (BL) represent the corresponding searching cones (see the text for further clarifications).}
    \label{fig:SkyMaps}
\end{figure}

After the collision, the probability distribution of the cosine of the scattering angle for the electron ($\mu_e$) in the observer's frame can be expressed as (more detailed derivations can be found in Ref.~\cite{wang2021direct})
\begin{equation}
    P(\mu_e;\,T_\chi) =\frac{2\mu_e\gamma_\CM^2(T_\chi)\Theta(1-\mu_e)}{\left[\mu_e^2+\gamma_\CM^2(T_\chi)(1-\mu_e^2)\right]^2}\,,
    \label{eq:Pe}
\end{equation}
where
\begin{equation}
\gamma_\CM^2(T_\chi) \equiv \frac{(T_\chi+m_\chi+m_e)^2}{\left(m_\chi+m_e\right)^2+2m_e T_\chi}
\label{eq:gammacm}
\end{equation}
is the squared Lorentz boost factor of the center-of-mass and the Heaviside theta function $\Theta$ ensures that $0\leq\mu_e\leq1$. From Eqs.~\eqref{eq:Pe} and \eqref{eq:gammacm} follow that the larger $T_\chi$ is, the more forward the motion of electron ($\mu_e\to1$) will be. For Bin1 and Bin2, a conservative half-opening angle of the searching cone ($\delta$) can be derived by imposing
\begin{equation}
    P\left(\mu_e>\cos \delta;\,T_\chi^\min(T_e = T^\min_l)\right) \gtrsim 0.95,
    \label{eq:cone}
\end{equation}
where $T_\chi^\min(T_e)$ is defined as in Eq.~\eqref{eq:Tpmin} with the substitutions $\chi \to e$ and $p\to\chi$, and $(l,\,T_l^\min/\text{GeV}) \in \{(\text{Bin1},0.1),\,(\text{Bin2}, 1.33)\}$. 
Note that $\gamma^2_\CM\left(T_\chi = T_\chi^\min(T_e)\right) = 1 + T_e/(2 m_e)$ and, consequently, Eq.~\eqref{eq:cone} is independent of $m_\chi$. 
For the third energy bin, given the limit of angular resolution of the detector, we simply set $\delta = 5^\circ$ \cite{Super-Kamiokande:2017dch}. 
The number of expected background in each cone ($N_\text{Bkg}^\delta$) can be estimated by assuming an isotropic distribution \cite{Super-Kamiokande:2017dch,PhysRevLett.122.181802}, while the corresponding number of data events inside each cone from TXS 0506+056 and BL Lacertae ($N_\text{TXS}^\delta$ and $N_\text{BL}^\delta$) can be counted directly. We depict in Fig.~\ref{fig:SkyMaps} the spatial distribution of data events that followed the data selection of Ref.~\cite{Super-Kamiokande:2017dch} together with the optimal searching cones around the two blazars.

Using the standard Poisson method \cite{Zyla:2020zbs}, we derived the 95\% Confidence Level (C.L.) upper limits on the event number from TXS 0506+056 ($N_\text{TXS}$) and BL Lacertae ($N_\text{BL}$). 
All the results of the data analysis have been condensed in Table~\ref{Tab:Super-K}. 
\begin{table}
\centering
\begin{tabular}{cccc}
\toprule
    \rowcolor[gray]{.95}
    \multicolumn{4}{c}{\bf Sensitivity of Super-Kamiokande}\\
    \hline
    \hline
     \rule{0pt}{2.5ex}  
    & $\text{Bin1}$ & $\text{Bin2}$& $\text{Bin3}$\\
    \hline
    \rule{0pt}{2.5ex}  
    $T_e$ (GeV) & $\left(0.1, ~1.33\right)$ & $\left(1.33, ~20\right)$ & $\left(20, ~10^3\right)$ \\
     $N_\text{Data}$   & 4042 & 658 & 3  \\
      $N_\text{Bkg}$   & 3992.9 & 772.6 & 7.4  \\
      \rule{0pt}{2.5ex}  
          $\epsilon_\text{sig}$   & $93.0\%$ & $91.3\%$ & $81.1\%$ \\
     $\delta$   & $24^\circ$ & $7^\circ$ & $5^\circ$  \\
     \rule{0pt}{2.5ex}  
      $N_\text{TXS}^{\delta}$  & 169 & 2 & 0  \\
      \rule{0pt}{2.5ex}  
       $N_\text{BL}^{\delta}$   & 167 & 4 & 0  \\
       \rule{0pt}{2.5ex}  
      $N_\text{Bkg}^{\delta}$   & 172.6 & 2.88 & 0.014  \\
      \rule{0pt}{2.5ex}  
        $N_\text{TXS}$ ($95\%$ C.L.)   & 19.39 & 3.42 & 2.98  \\
      $N_\text{BL}$ ($95\%$ C.L.)  & 17.27 & 6.27 & 2.98  \\
    \hline
    \hline
\end{tabular}
\caption{A summary of our data analysis that follows Ref.~\cite{Super-Kamiokande:2017dch}. 
Three bins are considered according to the electron recoil energy, namely Bin1, Bin2 and Bin3. 
For each bin, we adopt different half-opening angles ($\delta$) of the searching cone  based on Eq.~\eqref{eq:cone} and Super-K angular resolution. Using the Poisson method, the $95\%$ C.L. upper limits on the number of events from TXS 0506+056 ($N_\text{TXS}$) and BL Lacertae ($N_\text{BL}$) can be derived. See the text for further details.
}
\label{Tab:Super-K}
\end{table}

\subsection{Bounds on Scattering Cross Section}
\label{sec:resB}
The number of BBDM-induced electron recoil events at Super-K can be expressed as
\begin{equation}\label{eq:DM-N rate}
    N_e^\text{DM} \simeq  N_e\sigma_{\chi e} t_\text{obs}\int_{T_\text{exp}^\text{min}}^{T_\text{exp}^\text{max}}dT_e\, \int_{T_{\chi}^\min(T_e)}^{+\infty}\!\!\ 
  \frac{ dT_{\chi}}{T_{e}^{\mathrm{ max}}(T_\chi)}\frac{d\Phi_{\chi}^z}{dT_{\chi}}\,, 
\end{equation}
where $N_e = 7.5\times 10^{33}$ is the total number of electrons in the Super-K water reservoir \cite{Super-Kamiokande:2017dch}, $t_\text{obs} = 2628.1$ days is the exposure time, $\left[T_\text{exp}^\min,~T_\text{exp}^\max\right]$ is the energy range of each bin and $d\Phi_{\chi}^z/dT_{\chi}$ is the BBDM flux at detector. The corresponding constraints on $\sigma_{\chi e}$ from TXS 0506+056 (BL Lacertae) can be obtained by imposing 
\begin{equation}
    N_e^\text{DM} \times \epsilon_\text{sig} < N_\text{TXS}\, (N_\text{BL}).
    \label{eq:limit}
\end{equation}
    
For large enough $\sigma_{\chi p}$ and/or $\sigma_{\chi e}$, the DM flux will be attenuated by the scatterings with nuclei and/or electrons in the crust of the Earth and render a \virg{blind spot} at detectors \cite{Starkman:1990nj,Mack:2007xj,Hooper:2018bfw,Emken:2018run,wang2021direct}. If one ignores the form factor, the corresponding upper bound for $\sigma_{\chi p}$ ($\sigma_{\chi e}$) at XENON1T (Super-K) is about $3.0\times10^{-28}~\text{cm}^2$ ($2.0\times10^{-28}~\text{cm}^2$) \cite{wang2021direct,PhysRevLett.122.181802}. However, a recent study shows that, after the form factor is properly taken into account, the exclusion upper bound for $\sigma_{\chi p}$ can be increased at least by four orders of magnitude \cite{Xia:2021vbz}. Therefore, given such large values, we decide to ignore the upper constraints in our analysis. Concerning the lower exclusion limits, the Earth's attenuation effects can be safely neglected because of the extreme smallness of $\sigma_{\chi p}$ and $\sigma_{\chi e}$ ($\sigma_{\chi p}\lesssim 10^{-35}\,\text{cm}^2$ \cite{wang2021direct} and $\sigma_{\chi e}\lesssim10^{-33}\,\text{cm}^2$ \cite{Super-Kamiokande:2017dch}), meaning that we can safely replace $d\Phi_{\chi}^z/dT_{\chi}$ in Eq.~\eqref{eq:DM-N rate} with $d\Phi_{\chi}/dT_{\chi}$ given in Eq.~\eqref{eq:spectrumDM}.

For our results, we first fix $\sigma_{\chi p}$ at the corresponding lower exclusion values given in Ref.~\cite{wang2021direct} for BMP1 and BMP2, respectively. Therefore, there are only two free parameters, $\sigma_{\chi e}$ and $m_\chi$. 
Combining Eqs.~\eqref{eq:DM-N rate} and \eqref{eq:limit} we derive the $95\%$ C.L. limits on $\sigma_{\chi e}$ from Super-K results. 
However, it is worth noting that the limitations on $\sigma_{\chi e}$ we obtain in this case are the most optimistic results, because we have actually selected the maximal possible value of $\sigma_{\chi p}$. At the end of this section, we will discuss how different choices of $\sigma_{\chi p}$ would influence our final results.

\begin{table}
\centering
\begin{tabular}{ccccc} 
\toprule
\rowcolor[gray]{.95}
    \multicolumn{5}{c}{\bf Constraints on DM-electron Cross Section}\\
    \hline
    \hline
\rule{0pt}{2.5ex} 
\multirow{2}{*}{Source} & \multirow{2}{*}{$T_e$ (GeV)}~ & \multicolumn{3}{c}{$m_\chi$ (GeV)}  \\ 
\cline{3-5}
\rule{0pt}{3ex} 
                  &                   &
                  $10^{-6}$ & $10^{-4}$ &  $10^{-2}$ \\ 
\hline
\rule{0pt}{2.5ex} 
\multirow{3}{*}{} &   (0.1, 1.33)     &
$-38.35$ & $-38.05$ & $-37.39$  \\
BL Lacertae       &   (1.33, 20)      &
$-38.00$ & $-37.70$ & $-37.84$  \\
                  &   (20, $10^3$)    &
                  $-37.49$ & $-37.07$ & $-35.87$  \\
\hline
\rule{0pt}{2.5ex} 
\multirow{3}{*}{} &   (0.1, 1.33)     &
$-37.48$ & $-36.95$ & $-35.67$  \\
TXS 0506+056      &   (1.33, 20)      &
$-37.67$ & $-37.26$ & $-35.68$  \\
                  &   (20, $10^3$)    &
                  $-37.15$ & $-36.62$ & $-35.49$  \\
\cline{2-5}
\hline
\hline
\end{tabular}
\caption{The constraints on the DM-electron scattering cross section in the form $\text{log}_{10}[\sigma_{\chi e}/\text{cm}^2]$ for different source, energy bin and DM mass. Note that all these results are derived for the parameters of BMP1. Also, for $m_\chi/\text{GeV} = \{10^{-6},~10^{-4},~10^{-2}\}$, the corresponding lower boundary values of $\sigma_{\chi p}$ are $\text{log}_{10}[\sigma_{\chi p}/\text{cm}^2] = \{-34.80,\,-34.20,\,-32.29\}$ and  $\{-35.67,\,-35.36 -34.09\}$ for TXS 0506+056 and BL Lacertae, respectively.}
\label{Tab:res}
\end{table}

\begin{figure}
\centering
\includegraphics[width=0.48\textwidth]{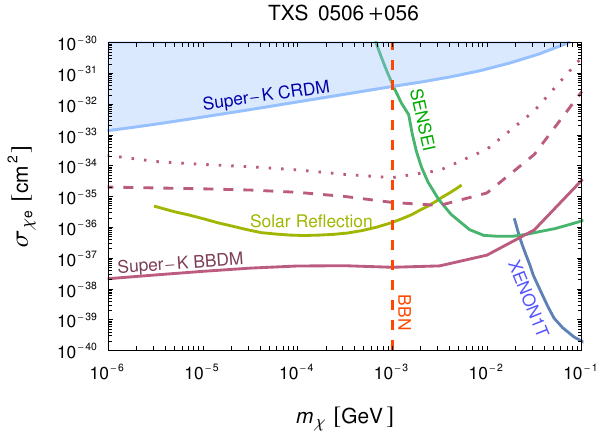}
\includegraphics[width=0.48\textwidth]{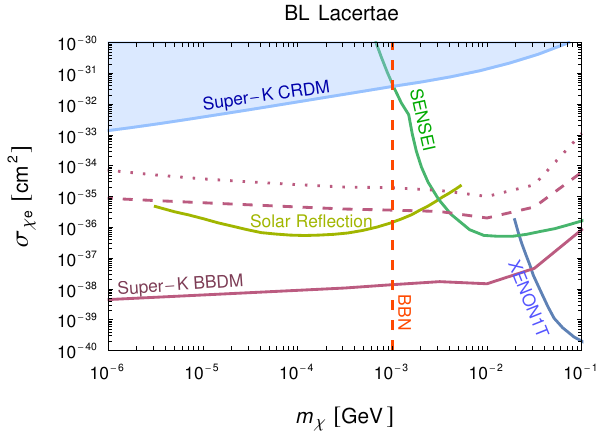}
\caption{The constraints on DM-electron scattering cross section imposed by Super-K \cite{Super-Kamiokande:2017dch}. The left panel is for TXS 0506+056, while the right panel for BL Lacertae. The solid, dashed, and dotted purple lines correspond to BMP1, BMP2, and BMP3, respectively. For comparison, the constrains from CRDM \cite{PhysRevLett.122.181802, PhysRevD.100.103011}, XENON1T \cite{XENON:2019gfn}, SENSEI \cite{SENSEI:2020dpa}, Solar Reflection \cite{PhysRevLett.120.141801}, and BBN \cite{Krnjaic:2019dzc, Depta_2019} are included.}
\label{fig:SEB}
\end{figure}

In Table~\ref{Tab:res}, we report the constraints on the logarithm of the DM-electron cross section ($\text{log}_{10}[\sigma_{\chi e}/\text{cm}^2]$) from the two considered blazars with BMP1 parameters, for different energy bins and various DM masses. We find that, for different $m_\chi$, the most stringent limits on $\sigma_{\chi e}$ may come from different bins. Hence, more stringent limits can be obtained by performing a combined analysis. We then show in the left and right panels of Fig.~\ref{fig:SEB} the results for TXS 0506+056 and BL Lacertae, respectively. The solid line corresponds to BMP1, while the dashed and dotted lines represent the results for BMP2 and BMP3, respectively. For each considered blazar, the differences between these lines arise only from $\Sigma_\DM^\text{tot}/m_\chi$ (see Fig.~\ref{fig:sigmaDM}). Compared to the results of CRDM, the BBDM constraints are more stringent by
orders of magnitude, depending on the DM mass and the parameters of the benchmark point.
Besides, the boundary values of $\sigma_{\chi e}$ we have obtained are much smaller than the ones adopted for $\sigma_{\chi p}$, meaning that the BBDM flux at Super-K is dominated by the contribution from the protons in the blazars' jets.

On the other hand, assuming crossing symmetry, one can relate the scattering cross section to that of the annihilation/production processes and consider the relevant bounds. For instance, the Big Bang Nucleosynthesis (BBN) could constrain the parameter space for DM masses less than $\mathcal{O}(1)$ MeV \cite{Krnjaic:2019dzc, Depta_2019}, although the precise bound is rather model dependent. From Fig.~\ref{fig:SEB} we find that, after including the BBN bound, the previous boundaries from, e.g., CRDM \cite{PhysRevLett.122.181802} and Solar Reflection \cite{PhysRevLett.120.141801}, are almost, if not all, excluded, whereas our results for BMP1 (as well as any intermediate case between BMP1 and BMP2) still cover an important region of the parameter space in the $1\lesssim m_\chi/\text{MeV} \lesssim 30$ range.

As we have emphasized in Sec.~\ref{sec:jet}, the luminosities of TXS 0506+056 and BL Lacertae vary over time. Therefore, during the entire exposure period of Super-K, both $L_p$ and $L_e$ are not constant. However, we estimate that smaller values for $L_p$ and $L_e$ would not affect too much our results.
For instance, since the boundary values of $\sigma_{\chi p} \propto L^{-1/2}_p$ \cite{wang2021direct} and $\sigma_{\chi e} \propto (L_p \sigma_{\chi p})^{-1} \propto L^{-1/2}_p$, if $L_p$ and $L_e$ are reduced by two orders of magnitude simultaneously, the lower exclusion limits for $\sigma_{\chi p}$ and $\sigma_{\chi e}$ would increase by a factor of 10.

\begin{figure}
\centering
\includegraphics[width=0.48\textwidth]{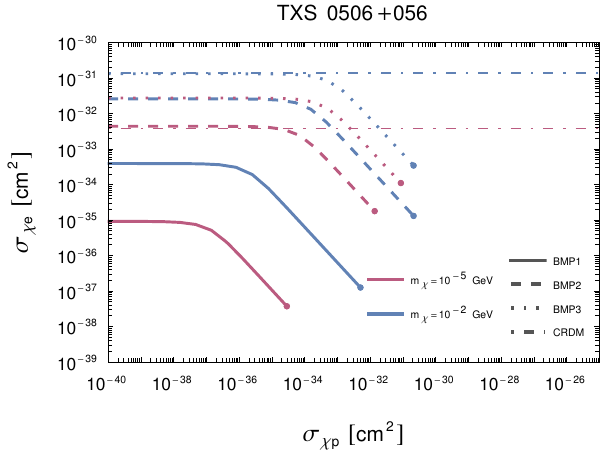}
\includegraphics[width=0.48\textwidth]{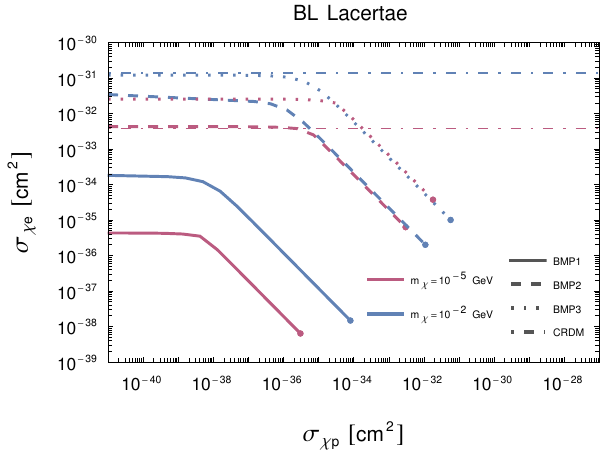}
\caption{The relationship between $\sigma_{\chi p}$ and $\sigma_{\chi e}$ for TXS 0506+056 (left panel) and BL Lacertae (right panel). Different colours correspond to different DM masses, namely $10^{-5}$ GeV (purple) and $10^{-2}$ GeV (blue). The solid, dashed, and dotted lines represent BMP1, BMP2, and BMP3, respectively. The dot-dashed horizontal lines indicate the lower boundary values of $\sigma_{\chi e}$ given by CRDM (see Fig.~\ref{fig:SEB}).}
\label{fig:xpxe}
\end{figure}

In Fig.~\ref{fig:xpxe} we show how the lower exclusion limit on $\sigma_{\chi e}$ changes for different values of $\sigma_{\chi p}$.
The left (right) panel corresponds to TXS 0506+056 (BL Lacertae). In the two plots, we show the curves for $m_\chi = 10^{-8}$ GeV (red), $10^{-5}$ GeV (green), and $10^{-2}$ GeV (blue), for both BMP1 (solid lines) and BMP2 (dashed lines).
We find that, for small values of $\sigma_{\chi p}$ (say $\sigma_{\chi p} \lesssim 10^{-38}~\text{cm}^{2}$), the exclusion boundaries of $\sigma_{\chi e}$ tend to constant values, which means that in this case the BBDM flux at Super-K would only come from the electron contribution, and can be regarded as the most conservative limits on the DM-electron cross section from BBDM. Note that the \virg{vanishing} $\sigma_{\chi p}$ scenario would be analogous to the case of purely leptonic SED model, for which the contribution from protons is naturally suppressed. For comparison, we indicate with horizontal dot-dashed lines the lower constraints on $\sigma_{\chi e}$ from CRDM (see Fig.~\ref{fig:SEB}).

%%%%%%%%%%%%%%%%%%%%%%%%%%%%
\section{Conclusion}
\label{sec:Conclusion}
The highly powerful jets emitted from the center, together with the large amount of DM present in their surroundings, make blazars ideal DM boosters. Based on Ref.~\cite{wang2021direct}, we have included the DM-electron interaction and derived the corresponding constraints on $\sigma_{\chi e}$ by making use of the available experimental results of Super-K.
In our analysis, two blazars have been considered, namely TXS 0506+056 and BL Lacertae.
Combining the spatial distribution of electron recoil data with the blazars' positions, we have conducted a refined analysis by setting a proper searching cone for different energy bins and derived the $95\%$ C.L. constraints on the DM-electron cross section with the standard Poisson method (see Fig.~\ref{fig:SEB}). 
Compared with the previous results from galactic CRs, the limits on $\sigma_{\chi e}$ from BBDM has improved by orders of magnitude, depending mainly on $m_\chi$ and the parameters relevant to the DM density profile. Besides, in view of future neutrino detectors such as Hyper-Kamiokande \cite{Hyper-K} and DUNE \cite{DUNE} (see also Ref.~\cite{Necib:2016aez}), our results could be further improved.

For future prospects, it could be interesting to do a more refined analysis (improving, e.g., the size of the searching cone and/or the time correlation) through the combination of neutrino detectors data with the observations of blazars from telescopes, such as Fermi-LAT \cite{Fermi-LAT} and/or the planned Cherenkov Telescope Array project \cite{CTA}. This, in principle, could enable us to select events from the right blazar at the optimal time, e.g.~during a blazar flare when the luminosity is enhanced, in analogy to the multi-messenger approach used to correlate the flaring of TXS 0506+056 with the first detection of cosmic neutrinos by IceCube Neutrino Observatory \cite{IceCube1, IceCube2,1807.04461, 1807.04748}. Furthermore, we expect that a statistical analysis extended to the full population of blazars would minimize the dependence on both model uncertainties and blazar selection, leading eventually to an enhancement of our results. Finally, studying possible BBDM signals
could allow to extrapolate more information on the SED jet models of blazars and/or the nature of DM.

%%%%%%%%%%%%%%%%%%%%%%%%%%%%
\begin{acknowledgments}
The authors wish to thank Serguey T.~Petcov for thoughtful discussions and suggestions. This work was supported by the research grant \virg{The Dark Universe: A Synergic Multi-messenger Approach} number 2017X7X85K under the program PRIN 2017 funded by the The Italian Ministry of Education, University and Research (MIUR), and by the European Union’s Horizon 2020 research and innovation program under the Marie Skłodowska-Curie grant agreement No 860881-HIDDeN.
\end{acknowledgments}

%\bibliography{BBDMSK}
%\bibliographystyle{JHEP}

\providecommand{\href}[2]{#2}\begingroup\raggedright\endgroup

\end{document}